# The Work Avatar Face-Off: Knowledge Worker Preferences for Realism in Meetings

Vrushank Phadnis*    Kristin Moore†    Mar Gonzalez-Franco‡

Google

## ABSTRACT

While avatars have grown in popularity in social settings, their use in the workplace is still debatable. We conducted a large-scale survey to evaluate knowledge worker sentiment towards avatars, particularly the effects of realism on their acceptability for work meetings. Our survey of 2509 knowledge workers from multiple countries rated five avatar styles for use by managers, known colleagues and unknown colleagues.

In all scenarios, participants favored higher realism, but fully realistic avatars were sometimes perceived as uncanny. Less realistic avatars were rated worse when interacting with an unknown colleague or manager, as compared to a known colleague. Avatar acceptability varied by country, with participants from the United States and South Korea rating avatars more favorably. We supplemented our quantitative findings with a thematic analysis of open-ended responses to provide a comprehensive understanding of factors influencing work avatar choices.

In conclusion, our results show that realism had a significant positive correlation with acceptability. Non-realistic avatars were seen as fun and playful, but only suitable for occasional use.

## 1 INTRODUCTION

Work practices have drastically evolved post pandemic, and so has the receptiveness of avatars [4, 33, 65]. Once popular only amongst gamers and virtual reality (VR) enthusiasts, avatars are now gaining adoption in other social and professional settings including remote meetings. Additionally, advancements in machine learning techniques have vastly improved 3D asset creation, democratized access to photorealistic avatars [1, 60]. And lastly, the recent upsurge of VR head mounted device (HMD) launches has revived interest in VR and in avatars [15].

Large deployments of avatars have already been done in VR, such as the Accenture on-boarding program during the pandemic, where employees met in digital twins of physical office spaces [66]. However, most of the research on avatars in professional settings has been restricted to lab experiments and case studies with small participant pools [53, 55]. Avatar acceptability is context-dependent, and their perception is still evolving [67]. The professional and social norms of representing oneself in virtual environments as an avatar are unclear [61].

Our work provides an assessment of avatar perception among knowledge workers from five countries using a large-scale survey. We study the acceptability of five levels of photorealism in avatars (See Fig. 1), in the new context of work meetings. In particular, we answer the following questions.

*e-mail: vrushank@google.com
†e-mail: kristinmoore@google.com
‡e-mail: margon@google.com

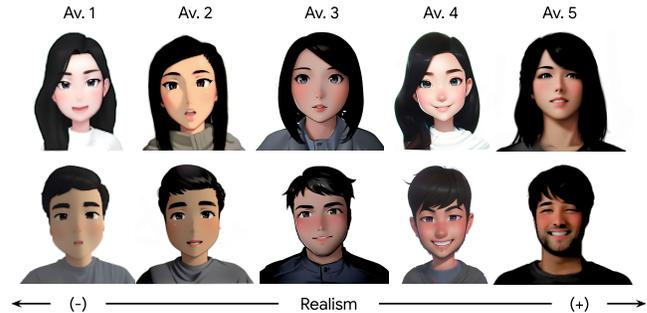

Figure 1: Representations of avatar assets used in the survey. Five male-female pairs were created using publicly available avatar systems. The pairs are arranged in an increasing order of realism from left to right. The difference in realism was found to be statistically significant.

**Primary research questions**

- **P1:** Does realism matter to knowledge workers when selecting a work avatar?
- **P2:** Does the acceptability of avatar realism level depend on the relationship with the avatar user?
- **P3:** Does the acceptability of avatar realism level vary by demographic and firmographic traits of knowledge workers?

**Secondary research questions**

- **S1:** What subjective attributes are most important when choosing an avatar for work meetings?

**Internal validity**

- **I1:** Is the difference in realism levels of the five avatar styles used in our survey statistically significant?

As seen in Fig. 5, survey respondents clearly favored realism in work avatars. Sect. 4.4 elaborates on this finding by visualizing the correlation between acceptance and realism ratings for each avatar type. Sect. 4.5 shows that expectations of avatar realism held true across all demographic attributes, including age, gender, and firmographic traits (Fig. 3), but differed across countries (Fig. 8). We also found that non-realistic avatars were rated differently (Sect. 4.3) based on familiarity to the impersonating work colleague.

Our evaluation of open-response comments in Sect. 4.6 suggests that knowledge workers value attributes like professionalism, credibility, and work-appropriateness when considering avatar styles for work. However, a few comments also suggest that less realistic avatars have a place in some contexts of work. They provide an avenue for fun and are perceived to be more charming.

As a secondary contribution of this paper, we document knowledge work behaviors like meeting schedules and remote work frequency in Table 1.

## 2 RELATED WORK

### 2.1 The use cases of avatars beyond VR

Avatar embodiment has long been popular in immersive environments accessed through VR HMDs such as Meta's Quest and Apple's Vision Pro [3, 59]. However, majority of avatar use predominantly happens on 2D interfaces using smartphone or web camera-based applications like Zoom video filters and Microsoft Teams' Mesh avatars [23, 32].

As VR HMDs have yet to gain mass adoption and smartphone/web-camera-based avatars are ubiquitous, 2D avatars continue to be an important area of research [13, 30, 45]. Examples of 2D avatars use include customized emojis, VTubing, and video filters [31, 64]. Typically, these use cases last for short duration and are context-specific, such as attending an online social event in virtual space for team bonding.

In some aspects, 2D avatars have a distinct advantage. For example, Liu et al. evaluated avatar-based garment try-outs in which participants found the smartphone-based AR avatars to be more realistic in their fit of a garment compared to the VR counterpart [43]. In another example, Li et al. reported that screen-based users experienced a lower mental workload while experiencing immersive content compared to HMD users [42].

Our work evaluates realism of avatars using 2D assets, and is builds upon prior research that has used screen-based avatars as a proxy for VR avatars [11, 29, 52]. In one such example, Zibrek et al. compared the likeability of avatars presented in VR versus video based avatars, across various realism levels [73]. Authors found that affinity scores were correlated with realism levels, but the medium (VR vs. video) did not play a significant role in avatar perception.

### 2.2 Measuring anthropomorphic perception of avatars

Anthropomorphism, or the human-like appearance of digital artifacts, has been studied in various domains of human-computer interaction (HCI) [5, 21, 50]. Researchers have explored realism levels to identify the ideal representation of humans in HCI applications [19, 26]. Although more realism is generally preferred, there is no clear consensus [44, 72]. Further, the resulting uncanny valley effects are difficult to overcome [18, 37, 40].

As coined by Mori et al., the uncanny valley effect states that the appeal of a digital representation of an inanimate object drastically drops as its form approaches, but fails to fully obtain human-like appearance [48]. This trade-off between realism and appeal has widely been studied in research settings, but large-scale field studies or surveys are uncommon [39].

Anthropomorphism has typically been measured using a Rasch scale-based questionnaire [10]. For example, McDonnell et al. used ratings that included "extremely abstract" to "extremely realistic" and "very unfriendly" to "very friendly", to measure the realism of avatars generated using various rendering methods [46]. We adopt McDonnell et al.'s rating system in our work to determine the realism of our avatar assets.

Avatar realism is perceived from a variety of attributes, such as appearance, facial expressions, body movement, and audio quality [25, 28, 57, 72]. It is also known that animation of the avatar has an effect on avatar perception and cognitive performance [25, 51]. Avatars that fully achieve human-likeness in all aspects are an active area of research, but not fully solved [69, 71].

Part of the challenge lies in accurately extracting the input parameters needed to drive avatar rigs [24]. For example, advanced photogrammetry setups involving multiple camera arrays and motion capture systems are prerequisites for creating high-fidelity renderings [9, 41]. Further post-processing and editing is often needed, making the overall process infeasible for real-time applications.

### 2.3 Avatars in the workplace

Although majority of avatar research in HCI is done in VR, it is mostly conducted in a laboratory settings and lacking in real-world scenarios [6, 16, 35]. Further, the limited availability of VR devices means that such work has been typically reserved for niche applications, like training and skill building [8, 27, 70].

On the contrary, the wider adoption of augmented reality (AR)-based avatars on smartphones and video has enabled researchers to access larger participant pools. Additionally, the ease of integrating 2D avatars in surveys means large scale data collection is possible and often used to create insights that are more generalized [7].

For example, Praetorius et al. surveyed 126 participants on avatar preferences for work, social, and gaming. Participants rated realism as necessary for meeting friends but not at work, contrasting our findings [54]. A noted limitation of this work is the disproportionate representation of younger aged participants in the survey panel.

Injpen et al. surveyed 1,020 participants to evaluate comfort in interacting with 31 avatar styles of various realism levels. Although the authors found support for more realistic avatars being preferred, it was difficult to isolate this finding as realism was not evaluated consistently across the avatar assets [36].

## 3 METHODS

### 3.1 Avatar assets

Five avatar pairs were adapted from publicly available avatar systems: Memoji, AREmoji, Horizon Worlds, MetaHumans, and ReadyPlayerMe. Avatars were created using a single person's facial image, emoting the same facial expressions; minimizing differences occurring from multiple inputs to the avatar systems. Avatar assets were shown in the form of GIF animations shown in our survey but only their representations are shown in Fig. 1, to preserve identity and user terms. The animations were tested in an initial pilot of 50 users to confirm fidelity of avatar realism.

Unlike prior work assessing avatar realism, we did not use any advanced puppeteering techniques to control our avatar renders [46, 56, 72]. We purposefully did not use a systematic approach in creating our assets because such a structured approach would be easily discernible by participants. This allowed us to qualitatively analyze attributes desirable in work avatars without fixating on specific aesthetic elements of the avatar visuals, such as lighting, shadowing, and shape, which are difficult to control outside of VR [68, 72]. We wanted our survey assets to be consistent with commercially available avatars which were already familiar to knowledge workers.

All participants rated the same five pairs of avatars, but in a randomized sequence, to limit ordering effects. Av. 1 was the least realistic, and Av. 5 approached photorealism. Throughout this work, we will refer to each avatar by their corresponding naming shown in Fig. 1.

### 3.2 Participants

2,509 participants were recruited equally across five countries: United States, South Korea, United Kingdom, Germany, and France. These countries were selected based on VR usage trends and to mitigate same-race bias, as our model avatars were Asian and Caucasian-looking [38]. Parity in gender, age groups, and geographic regions was maintained to create a representative participant pool. We screened participants based on their job profiles and their relevance to knowledge work [17, 63]. In addition, participants under the age of 18, or unemployed, or those who did not use video meetings for work were excluded.

Participants were reimbursed for their time, but there was no performance-based incentive.

### 3.3 Survey design

The survey was split into three subsections, as shown in Fig. 2. In the first section, participants answered demographic questions and were filtered based on an recruitment criteria. Participants provided information about their work habits in section two; such as meeting schedule, type of meeting attendees, amount of time spent working from home, and productivity tools used at work. The goal of this section was to capture any work meeting habits that could influence participants' avatar style choice. In the third section, participants rated each avatar style on realism and acceptance in work settings [46]. Participants took an average of 10 minutes and 30 seconds to complete the survey.

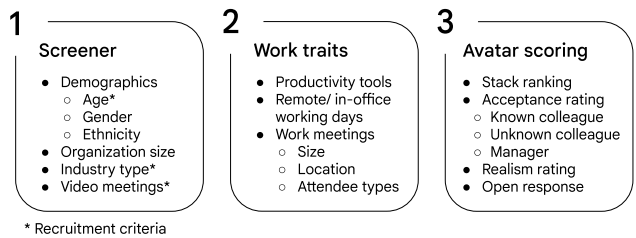

Figure 2: Survey flow showing each subsection with their corresponding question topics. Section 1 (leftmost) implemented a recruitment criteria to exclude participants that were under the age of 18 or were unemployed or did not use video meetings for work.

The survey was structured to measure within-subjects differences. It was administered in four languages: English, Korean, French, and German. Internal pilots were held with native speakers of each language for proofreading. Following which, a soft launch was done with 50 users in the United States before releasing to the entire survey pool. We partnered with an external agency for targeted recruitment of knowledge workers. With an incidence rate of less than 10%, the survey was fielded over a period of a month and a half to complete each user quota.

### 3.4 Data generation and analysis

Acceptability of avatar style was rated based on use-cases for scenarios with an unknown colleague, manager and known colleague. Each scenario was rated on a six-point Likert scale response from "Totally unacceptable" to "Totally acceptable." An overall acceptance score was calculated by aggregating scores across all scenarios for all avatars. Prompts used to assess acceptability were as follows:

Ⓤ An internal colleague that you haven't met before attends a project meeting with you as the below avatar style. Indicate how acceptable or unacceptable you feel it is for them to appear as this avatar style in this scenario:

Ⓜ Your manager attends a project meeting with you as the below avatar style. Indicate how acceptable or unacceptable you feel it is for them to appear as this avatar style in this scenario:

Ⓚ An internal colleague who you know well attends a project meeting with you as the below avatar style. Indicate how acceptable or unacceptable you feel it is for them to appear as this avatar style in this scenario:

Realism ratings for each avatar style were based on an anthropometric scale ranging from -100 to +100, corresponding to an avatar realism rating of "Extremely abstract" to "Extremely realistic" [46].

To determine our statistical modeling approach, we tested realism ratings and acceptance scores for normality using a Shapiro-Wilk test. Neither data set was found to be normal and determined our use of non-parametric tests. Throughout Sect. 4, we use Wilcoxon rank-sum tests to compare two means and Kruskal-Wallis tests with Dunn's post-hoc analysis (with Bonferroni adjustment) to compare three or more means.

To aggregate our numeric findings, an Elastic Net regression was formulated using acceptance scores as the dependent variable and demographics/knowledge worker traits as the independent variables. We initially used a grid search method to tune our model, but found a more optimal fit after manually iterating through the hyperparameters. The final version of the model used the hyperparameters of alpha = 0.1 and l1 ratio = 0.25, with a mean square error between 0.9 and 1.0. The model was built in two phases by iteratively excluding insignificant independent variables to reduce dimensionality.

In the open response section, participants responded to, "Please describe how you feel overall about this avatar style:".

### 3.5 Data validity

Various countermeasures were implemented to preserve data integrity. Thresholds on time-to-completion were used to check for speeding, straight-lining, and timeouts. Digital fingerprints were recorded and verified using GoIP cookies to discard duplicates and fraudulent responses. Randomized checking of the open-response comments was done manually during survey fielding.

## 4 RESULTS
### 4.1 Demographics and knowledge worker traits

Our participant pool had 1,275 males, 1,227 females, 4 non-binary, and 3 undisclosed genders. Age splits were as follows: 18-24 (808), 25-30 (294), 31-40 (557), 41-50 (492), 51-60 (270) and 61+ (88). Participants were equally distributed regionally with approximately 500 from each of the five countries.

A wide range of organization types and sizes were represented, as shown in Fig. 3. Employees of technology-focused companies were dominant in the survey pool, and organization sizes of 19 people and below were underrepresented.

266 participants were part-time employed, and the remaining 2,243 were full-time employed (30+ hours per week).

Table 1 shows information on meeting sizes, meeting attendee types, meeting location, and work schedules for our participant pool. This information, in addition to the demographics, will be evaluated for their influence on avatar acceptance scores in Sect. 4.5.

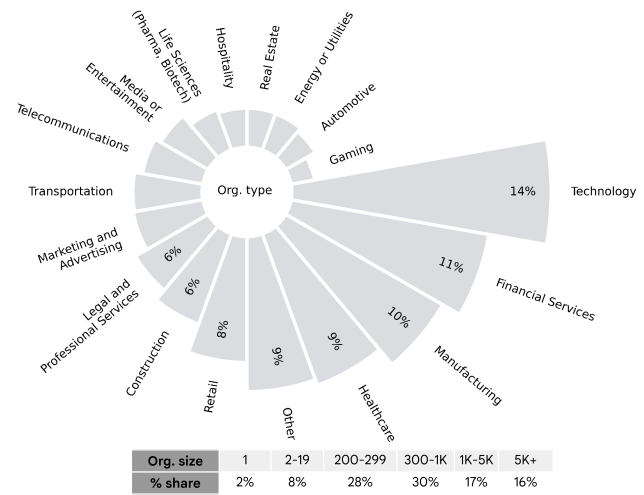

Figure 3: Organization types (top) and company sizes (bottom) of survey participants. Participants were equally spread out between organization types. Small and midsize enterprises with 200-1000 employees dominated the survey pool with 58% representation.

Table 1: Knowledge worker traits describing meeting characteristics and work schedule. Meeting size: Participants met in smaller groups (less than 5 participants) more frequently (multiple times a day or daily) and only occasionally met (less than once a week) in very large groups (more than 20 participants). Attendee location: Meetings with at least some remote attendees were more common than meetings with all in-office attendees. Attendee types: Participants met with their immediate internal team most often and only sometimes with external clients or collaborators. Remote work frequency: Most participants (846/2509) spent 1-2 days per week working from home.

| How often have you attended meetings of the following sizes in the last 30 days? | | | | | |
| --- | --- | --- | --- | --- | --- |
| | Never | Less than once a week | A few times a week | Daily | Multiple times a day |
| You and 1 additional attendee | 172 | 512 | 966 | 580 | 279 |
| Small meeting (2-5 participants) | 88 | 463 | 1182 | 564 | 212 |
| Mid-size meeting (6-10 participants) | 133 | 746 | 1118 | 356 | 156 |
| Large meeting (11-20 participants) | 260 | 1087 | 736 | 290 | 136 |
| Very large meeting (more than 20 participants) | 470 | 1209 | 480 | 470 | 114 |

| How often have you attended each of the following meeting types in the last 30 days? | | | | | |
| --- | --- | --- | --- | --- | --- |
| | Never | Less than once a week | A few times a week | Daily | Multiple times a day |
| Fully remote attendees | 208 | 653 | 1072 | 378 | 198 |
| Mostly remote / some in-office | 212 | 610 | 1114 | 401 | 172 |
| Mostly in-office / some remote | 198 | 644 | 1001 | 468 | 180 |
| Fully in-office attendees (multiple offices) | 349 | 660 | 869 | 441 | 190 |
| Fully in-office attendees (same office) | 353 | 643 | 838 | 477 | 198 |

| How often have you met with the following individuals in the last 30 days? | | | | |
| --- | --- | --- | --- | --- |
| | Never | Rarely | Sometimes | Often |
| Immediate internal team members (e.g. on a project team, in similar roles) | 82 | 329 | 990 | 1108 |
| Extended internal team members (e.g. within a program area, across program areas) | 117 | 441 | 1194 | 757 |
| Extended internal colleagues (e.g. co-workers across the organization) | 145 | 515 | 1205 | 644 |
| External clients / customers (i.e. individuals not in your organization) | 297 | 632 | 1043 | 537 |
| External partners / collaborators (i.e. individuals not in your organization) | 302 | 672 | 1074 | 461 |
| Other | 970 | 594 | 699 | 246 |

| On an average work week, how many days do you work from home? | | How often do you use video calling for work-related tasks? | |
| --- | --- | --- | --- |
| 5 days per week | 418 | Often throughout the day | 287 |
| 3-4 days per week | 553 | A few times a day | 546 |
| 1-2 days per week | 846 | Daily | 579 |
| Rarely/never work from home | 580 | A few times a week | 1097 |
| No typical schedule | 112 | | |

### 4.2 Realism ratings (internal validity)

As shown in Fig. 4, there was a noticeable difference in the realism values of the five avatars. The difference in means was found to be statistically significant (H(4) = 2991.6, p < 0.05), and all within-group differences were also found to be significant in post-hoc analysis.

### 4.3 Avatar acceptability

The average acceptability score across all avatar styles and scenarios was 3.9, which maps to an "Acceptable" rating on our Likert scale. Fig. 5 was created by aggregating the acceptance scores for each scenario mentioned in Sect. 3.4 and plotting the corresponding percentage splits. All within-group differences were found to be significant (H(4) = 1337.9, p < 0.05), except between Av. 3 and Av. 4. Av. 5 was given the highest acceptability rating and Av. 1 was rated the lowest.

Within each avatar style, acceptability scores for unknown colleagues were worst and scores for known colleagues were the best. These differences were found to be statistically significant. However, for Av. 5, all three work colleague scenarios were rated similarly.

### 4.4 Avatar acceptance vs. realism

The binned histogram in Fig. 6 shows the relationship between realism scores and acceptance scores. All avatars exhibited a positive correlation between realism and acceptance scores.

A Spearman's correlation model assessing the relationship between realism and acceptance scores of all avatars together was significant (rho = 0.32, p < 0.05), but with a low rho value, indicating a weak relationship. Similar results were found for individual correlations for each avatar (see Fig. 6 legend), except for Av. 1, which had a strong relationship. All results showed a monotonically increasing relationship between realism values and acceptance scores.

### 4.5 Regression analysis with avatar acceptability as the dependent variable

The top three positive and negative coefficients from the regression model are shown in Table 2. The model was built with acceptance scores as the dependent variable and demographics/knowledge worker traits as the independent variables. Independent variables relating to country, also referred to as regional or geographic differences in this paper, had a higher effect on acceptance scores than other variables, followed by work-from-home schedules and organization types. Acceptance score differences across gender, age, and the remaining knowledge worker traits were not significant.

#### 4.5.1 Independent variables with significant effect

Expanding on the Elastic Net regression results, Fig. 7 shows acceptance ratings of the five avatars divided across countries of origin, organization types, and working schedules. The average acceptance rating was 59.2, but participants from the United States and South

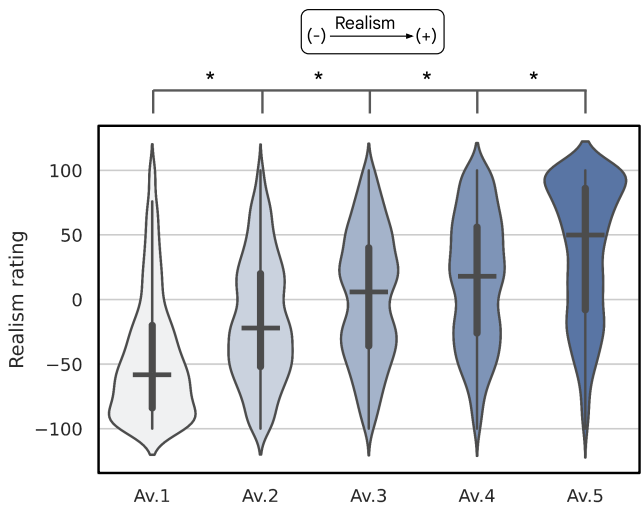

Figure 4: Aggregate realism scores for each avatar style. Av. 1 was rated as least realistic, while Av. 5 was rated as most realistic. Av.3 and Av. 4 were rated least apart in terms of realism. All avatar realism differences were found to be statistically significant.

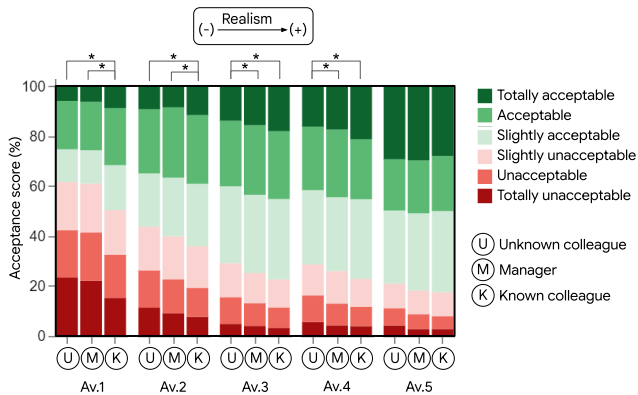

Figure 5: Aggregate acceptance scores for avatar use by unknown colleague, manager and known colleague. In line with realism scores, acceptability scores trended from low to high for Av. 1 to Av. 5.

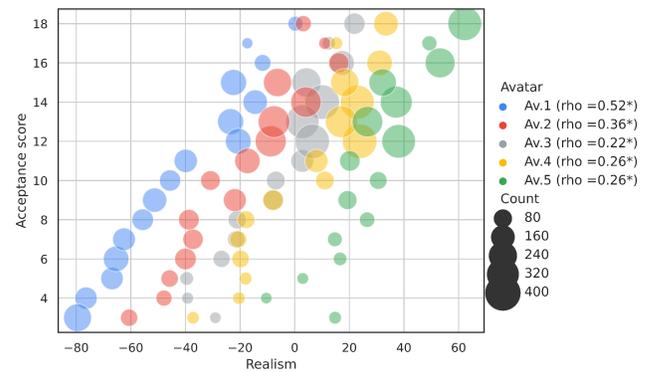

Figure 6: Binned scatter plot showing relationship between realism and acceptance scores for all avatar styles. Acceptance score (range: 0-18) shown in this figure is the sum of scores for a given avatar across all three work colleague scenarios.

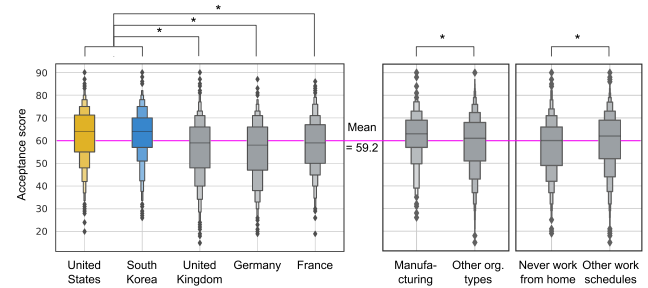

Figure 7: Avatar acceptance scores for (left) all five countries, (middle) manufacturing vs. non-manufacturing organization types and (right) respondents never working from home vs. other work schedules

Korea rated the avatars noticeably higher than the mean and are colored differently for emphasis. All differences within-group for countries were found to be significant (H(4) = 132.2, p < 0.05). However, in post-hoc analysis, differences between South Korea and United States participant responses were not found to be significantly different.

The difference between manufacturing and non-manufacturing organization types was not visually identifiable but statistically significant (Z = 3.9, p < 0.05). Similarly, the difference in acceptance scores of respondents who never worked from home and others was not visually identifiable but statistically significant (Z = -3.9, p < 0.05).

### 4.5.2 Avatar acceptance per country

To elaborate on our findings from Sect. 4.5.1, avatar acceptance scores split across countries are shown in Fig. 8. All group differences were found to be significant: Av. 1 (Z = 244.1, p < 0.05), Av. 2 (Z = 92.6, p < 0.05), Av. 3 (Z = 50.3, p < 0.05), Av. 4 (Z = 71.1, p < 0.05) and Av. 5 (Z = 57.2, p < 0.05). However, post-hoc analysis revealed varying results, denoted by asterisks (*) on the respective plots in Fig. 8. Av. 1 and Av. 4 showed significant differences between participants from the United States and South Korea versus other countries. However, the remaining avatars did not exhibit this difference.

### 4.6 Open response analysis

Given the scale of our dataset (12,545 comments), we used a combination of automated and manual analysis for full coverage. The automated technique helped us build an intuition of the overall opinion of work avatars, and the manual analysis provided an in-depth understanding of participants' rationale for work avatar selection.

#### 4.6.1 Sentiment analysis

Valence scores were calculated using VADER (Valence Aware Dictionary for Sentiment Reasoning), a module of NLTK (Natural Lan-

Table 2: Coefficients and z-scores of the regression analysis with acceptability ratings as the dependent variable. Country of residence had the strongest effect on avatar acceptability from all the independent variables. In the table, WFH: work from home.

| Independent variable | Acceptance (z-score) |
| --- | --- |
| United States (country) | 1.3 (0.10) |
| South Korea (country) | 1.3 (0.10) |
| Manufacturing (org. type) | 0.5 (0.04) |
| United Kingdom (country) | -0.5 (-0.04) |
| Never WFH (WFH) | -0.6 (-0.05) |
| Germany (country) | -0.9 (-0.07) |

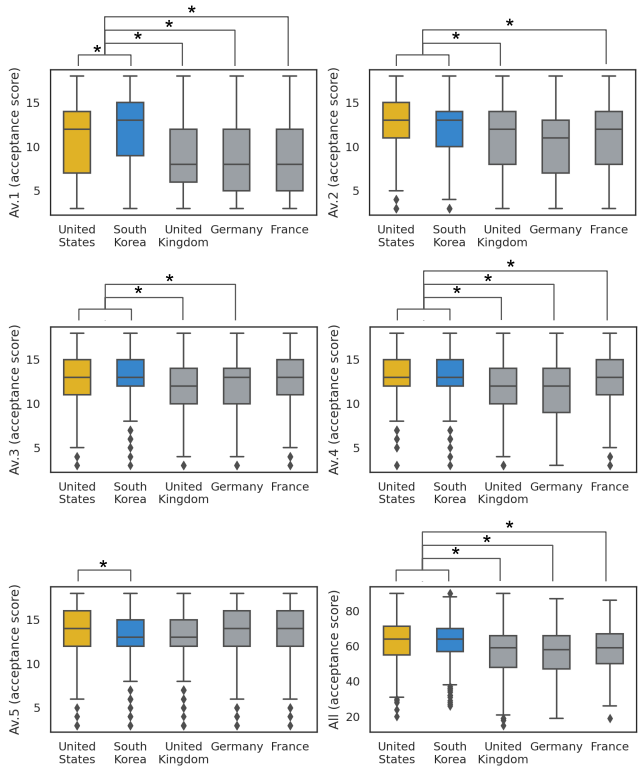

Figure 8: Avatar acceptance scores per country. Acceptance score (range: 0-18) is the sum of scores for a given avatar across all three work colleague scenarios. And Acceptance score (range: 0-90, shown in bottom right) is the sum of avatar scores across all five avatar styles.

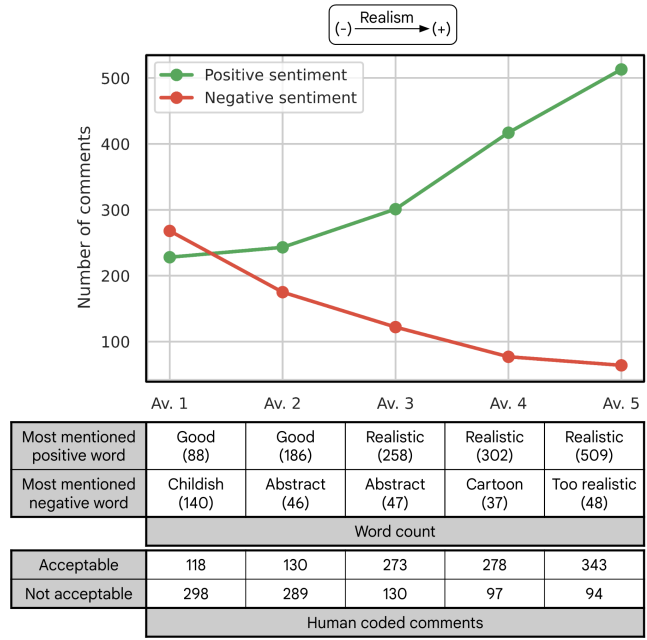

Figure 9: (Top) Sentiment analysis valence scores, (middle) corresponding most frequent words and (bottom) summary of manually coded scores

guage Toolkit). Fig. 9 shows the number of positive and negative comments for each avatar. The data is plotted as a line plot to highlight trends; with avatars listed in the order of realism ratings on the x-axis. Av. 1 was the only avatar that received more negative comments than positive. Av. 5 garnered the most positive comments and also showed the biggest difference between positive and negative scores. The most frequently uttered words remained somewhat consistent across avatars, but varied in frequency counts.

To validate the sentiment analysis results, 2,357 comments were randomly selected for grading by human coders. Comments were categorized as "Acceptable", "Not acceptable", and "Neither acceptable nor unacceptable". The coding was done by two researchers with 73.8% agreement, showing a high inter-rater reliability. More comments supported the acceptance of avatars (1,142) than not (908). The overall spread of human-coded comments, as shown in the bottom of Fig. 9, agreed with the sentiment analysis results.

### 4.6.2 Thematic analysis

The top 500 comments (100 per avatar) were selected based on length for an inductive thematic analysis [14]. A summary of the results is shown in Table 3. In the first category, 70 comments were categorized as *Real and acceptable*. 28 participants found the human-like appearance of realistic avatars desirable. A participant from South Korea described Av. 5 as *"It's realistic, so I feel like I'm talking to a real person, and... I will feel comfortable."* Another example from the United Kingdom, described Av. 5 as, *"It looks extremely realistic and I would not mind if my colleges or manager attended the meeting with this avatar as it looks exactly like them"*. 28 participants described photorealism as a necessity to portray a serious and professional outlook at work. A quote from a French participant regarding Av. 5 illustrates this: *"Very good avatar quality. This is very important because we are not in social networks"*. And 11 participants rated realism was required to build trust and authenticity in work meetings. As an example, two quotes from Germany about Av. 5 were as follows. *"The world is not yet anonymous enough that nobody can show his face, the only acceptable variant"* and *"Very realistic and trustworthy. Also appropriate because the person is easy to recognize"*.

141 comments were categorized under *Not real, but acceptable*. 54 participants rated non-realistic avatars as striking the right balance by being appealing and not eery. For example, a United States participant said, *"Not really realistic, but not offensive. Better to look like an avatar that is more general like these"* about Av. 2. And a participant from Germany rated Av. 4 as *"I find it pleasant because it is a mixture of real and comic, and yet it does not seem dubious"*. 18 participants favored non-realistic avatars because they were playful and friendly. *"It is a friendly character. You will be able to express your opinion more actively"* was said for Av. 1 by a participant from South Korea. And *"Unrealistic avatars that can serve well for a playful embodied in a conference"* was said for Av. 3 by a German participant. 27 participants appreciated the simplicity of non-realistic avatars. For example, A participant from the United Kingdom said, *"It's okay, it's cartoony but you can understand the general features of the person"* about Av. 4. Lastly, 25 participants preferred non-realistic avatars for ad-hoc or occasional use cases. For example, a German participant said, *"I find it very lively and cartoon, but at the same time friendly for use on special occasions"* for Av. 4.

173 comments were categorized as *Not real, not acceptable*. Out of which 18 tagged the non-real avatars as being childish. A participant from the United States said, *"This avatar style looks childish I wouldn't want to be seen like this unless I was a child myself"* regarding Av. 1. Some participants (N=37) associated the non-realistic avatars with video games/social media and not appropriate for work. For example, Av. 3 was rated as, *"I do not like. Looks like cartoons... Not possible to have avatars like those during our*

Table 3: Thematic analysis categories of the 100 longest comments for each avatar (501 total) split across all five countries. In the table, US: United States, SK: South Korea, UK: United Kingdom, DE: Germany, FR: France

|  | US | SK | UK | DE | FR | Total |
|---|---|---|---|---|---|---|
| *Realistic and acceptable* | | | | | | |
| Human-like/relatable | 5 | 5 | 10 | 8 | 0 | 28 |
| Work appropriate | 5 | 1 | 10 | 5 | 7 | 28 |
| Recognizable/trustworthy | 2 | 2 | 2 | 4 | 1 | 11 |
| *Non-realistic but acceptable* | | | | | | |
| Providing ideal trade-off | 13 | 7 | 14 | 13 | 6 | 59 |
| Simplistic representation | 3 | 9 | 9 | 4 | 2 | 27 |
| Good for occasional use | 6 | 3 | 5 | 8 | 3 | 25 |
| Friendly and playful | 3 | 5 | 4 | 5 | 1 | 18 |
| *Non-realistic and not acceptable* | | | | | | |
| Unprofessional/offensive | 12 | 5 | 19 | 5 | 12 | 53 |
| Fake/abstract appearance | 5 | 20 | 13 | 10 | 2 | 50 |
| Associate with gaming | 6 | 4 | 13 | 6 | 8 | 37 |
| Considered childish | 4 | 1 | 3 | 5 | 5 | 18 |
| *Realistic but not acceptable* | | | | | | |
| False impersonation | 3 | 1 | 6 | 1 | 2 | 13 |
| Uncanny feeling | 2 | 4 | 3 | 1 | 2 | 12 |

*meetings"* by a French participant. A participant from the United Kingdom rated Av. 2 as *"These avatars feel a little too generic - like from a 90s videogame. Ok - but not the best out of the bunch offered"*. 50 participants rated the non-real avatar as abstract looking. For example, a French participant rated Av. 1 as, *"This avatar is not at all done... Too far from reality. It does not give off any seriousness in a company"*. 53 participant rated non-realistic avatars as non-professional and offensive. A German participant said, *"I find it ridiculous, extremely ridiculous. If an employee uses such an avatar, I would pronounce the immediate termination due to disrespectful"* about Av. 2. And a French participant said, *"The features are far too much in a cartoon version and does not bring credibility to be used in the professional environment"* regarding Av. 1.

31 participants rated realistic avatars as unacceptable. 12 of them rated them as uncanny. A participant from South Korea described Av. 5 as, *"It's too realistic, so there's a bit of a sense of heterogeneity. The impression is a bit awkward and unnatural"*. And a French participant said, *"Extremely realistic they are almost no more avatars if they are modeled on the person who communicates. They could suddenly be a little disturbing in the professional context ..."* about Av. 5. 13 participants were concerned about possible misrepresentation using someone else's realistic avatars.

## 5 DISCUSSION

In this section, we summarize our observations, identify research extension opportunities, and contrast our findings with previous work. Before answering the primary research questions, we confirmed the range of realism ratings provided by our avatar assets (Fig. 4). This was an important validity check (I1), as insignificant differences would mean that our study assets could not provide enough resolution to test the primary research questions (P1/P2/P3).

### 5.1 Knowledge workers prefer photorealistic avatars for work meetings, but not in all situations (P1)

In Fig. 5, Av. 5 was rated as the most acceptable and Av. 1 as the least. Av. 2, Av. 3, and Av. 4 acceptability scores followed the same order as their realism scores. The correlation analysis in Fig. 6 confirmed this, wherein the baseline realism (x-intercept) of all the avatar scores matched findings from Fig. 4 and Fig. 5. Av. 5 has the highest baseline realism rating and Av. 1 has the lowest.

Overall, our results agree with the general consensus from prior work on avatars: more realism is favorable [20, 58]. However, prior findings are majorly tested in social settings and do not address the nuances of work avatars. Our results addresses this gap by contextualizing the effects of avatar realism for work use. Although we found that acceptability of work avatars varied linearly with realism levels, we also note the uncanny effects of the most realistic avatars. Some participants found that less realistic avatars strike the right balance between fun and eeriness.

### 5.2 Avatar realism is more important when meeting unknown colleagues vs. known (P2)

Avatars were perceived as more acceptable when meeting with known colleagues than when meeting with managers. And least acceptable to meet with an unknown colleague. This hierarchy was more pronounced for less realistic avatars. As noted by the '*' brackets in Fig. 5, known-unknown and known-manager differences were significant for the less realistic avatars: Av. 1 and Av. 2. On the other hand, known-unknown and unknown-manager differences were significant for Av. 3 and Av. 4. None of the Av. 5 scenarios were significantly different ($H(3) = 2.4$, $p = 0.31$).

We found that avatar acceptance for various work relationships depends on realism and proximity of the work relationship. Less realistic avatars are more acceptable when meeting with known colleagues. This finding is in agreement with Zibrek et al. [74], wherein the authors found that avatar realism goes beyond its stylization and is a factor of the impersonator's personality as well. As realism increases, it becomes more acceptable to meet with others that you know; but still perceived as less acceptable for unknown colleagues. Until you reach photorealism. At this point avatars are considered acceptable regardless of the relationship to the person. This suggests a strong preference for realistic work avatars across all settings and any accommodation for less realism is directly related to the closeness of work relationships.

### 5.3 Country of residence influenced avatar choices but firmographic and demographic factors did not (P3)

Avatar acceptability varied the most based on geographic regions, as seen in Fig. 7. Respondents from the United States and South Korea were more accepting of avatars for work meetings. And avatar acceptability did not significantly vary by age, gender, or employment characteristics. Nor did avatar acceptability vary based on firmographic information like meeting frequency, work from home schedule, and organization type/size. This could be because of cultural and work perceptions in different parts of the world [47]. For example, in Table 3 non-realistic avatars were rated as unprofessional and offensive by more participants in France, Germany, and the United Kingdom than elsewhere. The same subgroup of countries considered photorealistic avatars as human-like and work appropriate.

We also report a statistically significant difference between avatar acceptability ratings for participants from manufacturing vs. non-manufacturing organizations and for participants that never work from home vs. other work schedules. Given the size of these differences, as compared to the country related differences, they were not pursued for further analysis.

The most realistic avatar, Av. 5, was favored equally by participants of all countries as shown in Fig. 8. However, the remaining avatars were rated higher by participants from the United States and South Korea than others. This was specially true for Av. 1 and Av. 4. It seems that certain stylistic avatar choices are more appealing in some countries. Further work is needed to investigate the mapping of avatar stylizing to cultural subtleties in different countries.

### 5.4 Avatar choices were dictated by trustworthiness, professionalism, playfulness (S1)

The sentiment analysis results (Fig. 9) conforms to the acceptance score trends: avatars with higher acceptance scores had more positive sentiment scores, and vice versa. Av. 5 received the most positive comments and was rated as being professional and human-like. This is in line with Zibrek et al.'s findings, which found that a higher level of photorealism facilitates subtle human expressions, allowing more information to be perceived [75]. As described by a United Kingdom participant (for Av. 5): *"I probably would use this! Deffo the best out of all the avatars! The detail is amazing! Perfect for when I don't look put together and still want to be on a meeting!"* This finding is in accordance with Billieux et al., who found that avatars provide an escape from our daily lives by portraying an alternative, more appealing version of ourselves [12].

However, there could be work contexts where an abstract avatar can be advantageous. Non-realistic avatars were perceived as providing an ideal balance. For example, a United States participant said of Av. 2: *"The avatars are about 50/50 when it comes to being abstract or realistic. I still find them acceptable"*. Non-realistic avatars can also offer an opportunity for fun and better expressions. Respondents appreciated the lively/cute nature of the less realistic avatars. At the same time, they were called out for being unprofessional and not suitable for work. A German participant's quote for Av. 1 illustrates this: *"I actually find it totally cute, but for the meeting and the world of work + bosses probably not so well, since the two of them are not taken seriously"*.

On the other extreme, most realistic avatars suffered from the uncanny valley effect, where some participants thought that they looked unnatural or awkward. This leads us to believe that an optimal avatar for knowledge work offers the candidness of cartoon-like avatars without compromising on the professionalism needs of knowledge worker meetings. As an example, a United Kingdom participant said of Av. 3: *"..I feel this one would be accepted by most in a meeting, doesn't seem very unprofessional and also not so realistic that it makes others uncomfortable"* by a United Kingdom participant.

### 5.5 Knowledge worker traits

As a secondary contribution, our work documented various post-pandemic knowledge worker traits for archival knowledge. Although they did not prove to directly influence avatar choices in our work, these findings might prove useful for other researchers to build upon. As shown in Table 1, participants met in smaller meeting groups (2-5 and 6-10 attendees) and with their immediate internal team the most often. Most participants worked from home two days a week and used video calling infrequently (a few times a week).

### 6 LIMITATIONS

Our work found that differences between countries mattered the most, emphasizing the need for testing with a wider range of avatar and participant ethnicity. Although our survey pool had representative demographic backgrounds within each country, it would be beneficial to expand beyond the five countries, especially in countries where technology penetration is lower [49].

Our research relied on survey takers' assessment of hypothetical future work scenarios. Further, respondents did not have control over the avatar animations shown to them. This limits the applicability of our work to use cases in VR, given that human motion is integral to avatar perception [34]. Although our work discusses the implications of avatars by meeting attendees, we do not consider all workplace meeting contexts [2]. To fully ascertain the relevance of our results in professional settings, a longitudinal study tracking avatar choices across various work meeting contexts is a possible next step.

A survey-based approach limited us from presenting the avatar systems in an immersive environment like VR. Our results provide a starting point for such embodiment, but need follow-up work to test other nuances to realism like full-body avatar movement and voice.

Lastly, we did not test self-impersonation and also did not evaluate multi-user (collaborative) use cases. The implications for avatars embodied by others vs. self are different; potentially with higher uncanny valley effects for self-impersonation [22, 62].

### 7 CONCLUSION

Previous studies either surveyed a single geographic region or evaluated avatar realism in a lab setting. In contrast, our broad survey coverage enabled us to compare different demographic groups and provide a more comprehensive understanding of knowledge workers' needs and preferences for work avatars. Although we found significant empirical evidence supporting the use of realistic avatars in work settings, our results also highlight the importance of considering socio-cultural aspects when designing work avatars. Our findings provide valuable insights for researchers, developers, and workplace administrators to better position avatar deployments in the workplace.

Our survey panel included 2509 knowledge workers from a diverse set of countries, organizations and demographics. The difference in realism of avatar assets used in the survey was statistically significant, providing internal validity of our method. The main contributions of our survey are as follows:

Firstly, we found that the realism of avatars had a significant impact on their acceptability for work meetings. Avatars that were less realistic were rated as more acceptable for use by known colleagues than by managers and unknown colleagues. The most realistic avatar was rated as equally acceptable for all three work scenarios.

Secondly, the acceptability of avatars varied by country of origin. Participants from the United States and South Korea rated all avatars positively, while participants from other countries rated the less realistic avatars (Av.1, Av.2 and Av. 3) less positively. Other demographic and knowledge work factors did not have a significant effect on avatar choices.

Lastly, in our analysis of open responses, the most commonly mentioned work avatar attributes were professionalism, credibility, and seriousness. Participants also mentioned cuteness, fun, and liveliness as an additional benefit of using avatars at work.

As a secondary archival contribution, we document knowledge work trends like working two days from home, meeting more often with known colleagues and meeting in smaller groups (2 members or less).

### ACKNOWLEDGMENTS

We thank our anonymous reviewers and fellow Google colleagues for their insightful feedback. We would also like to extend our thanks to Barak Moshe for providing the assets for the survey.